\documentclass[pra,onecolumn,showpacs,nobalancelastpage,tightenlines,floatfix,amsfonts,amssymb]{revtex4}
\usepackage{graphicx}

\begin{document}

\title{Theory of Photon Blockade by an Optical Cavity with One Trapped Atom}
\author{K.~M.~Birnbaum}
\author{A.~Boca} 
\author{R.~Miller}
\author{A.~D.~Boozer}
\author{T.~E.~Northup}
\author{H.~J.~Kimble}
\affiliation{Norman Bridge Laboratory of Physics 12-33\\
California Institute of Technology, Pasadena, CA 91125}
\date{July 6, 2005}

\begin{abstract}
In our recent paper \cite{birnbaum05}, we reported observations of photon
blockade by one atom strongly coupled to an optical cavity. In support of
these measurements, here we provide an expanded discussion of the general
phenomenology of photon blockade as well as of the theoretical model and
results that were presented in Ref.~\cite{birnbaum05}.  We describe the general condition for photon blockade in terms of the transmission coefficients for photon number states.  For the atom-cavity system of Ref.~\cite{birnbaum05}, we present the model Hamiltonian and examine the relationship of the eigenvalues to the predicted intensity correlation function.  We explore the effect of different driving mechanisms on  the photon statistics.  We also present additional corrections to the model to describe cavity birefringence and ac-Stark shifts.
\end{abstract}

\maketitle

\section{Introduction}

The phenomenon of photon blockade, first proposed in Ref.~\cite{imamoglu97} in analogy
with Coulomb blockade for electrons \cite{fulton87,kastner92,likharev99}, occurs when the absorption of a
first input photon by an optical device blocks the transmission of a
second one, thereby leading to nonclassical output photon statistics.
Photon blockade has been predicted in many different settings \cite{grangier98,werner99,rebic99,rebic02,kim99,smolyaninov02},
including for a single two-level atom in cavity QED \cite{rebic02,tian92,brecha99,hood-thesis}. In the
latter setting, the blockade is due to the anharmonicity of the Jaynes-Cummings
ladder of eigenstates \cite{jaynes63}. If an incoming photon resonantly excites the
atom-cavity system from its ground state to $|1,\pm\rangle$ (where $|n,+(-)\rangle$ denotes
the $n$-excitation dressed state with higher (lower) energy), then a second photon
at the same frequency will be detuned from either of the next steps up
the ladder, i.e. from states $|2,\pm\rangle$. In the strong coupling regime \cite{kimble98},
for which the coherent rate of evolution $g_0$ exceeds the dissipative rates
$\kappa$ and $\gamma$, this detuning will be much larger than the excited-state line widths,
so that the two-excitation manifold will rarely be populated. This in turn
leads to the ordered flow of photons in the transmitted field, which
emerge from the cavity one at a time.

\begin{figure}
\includegraphics[width=5in]{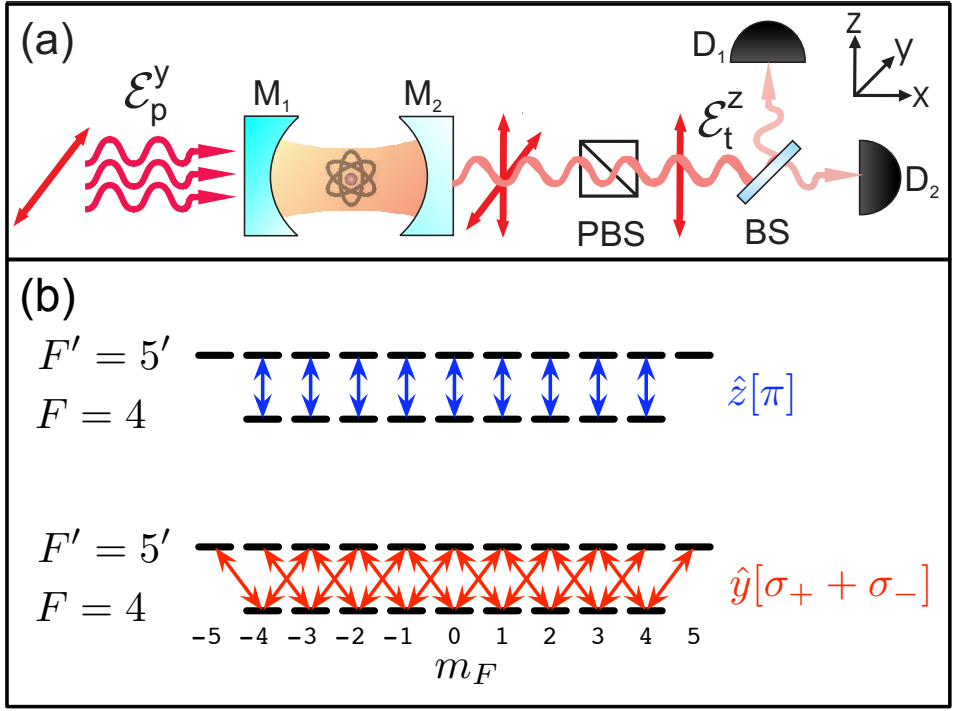}
\caption{(a) Diagram of the experiment of Ref.~\cite{birnbaum05}; (b) illustration of the dipole operators corresponding to the two linearly polarized cavity modes.  Here the atomic transition is $6S_{1/2},F=4\leftrightarrow 6P_{3/2},F^{\prime }=5$ in Cesium.}
\label{system}
\end{figure}

We recently reported observations of
photon blockade in the light transmitted by an optical cavity
containing one atom strongly coupled to the cavity field
\cite{birnbaum05}. For coherent excitation at the cavity input, the
photon statistics for the cavity output displayed both photon
antibunching and sub-Poissonian photon statistics. However, as
illustrated in Fig.~\ref{system}, the multiplicity of atomic and cavity states makes our experiment considerably
more complex than the simple situation described by the extended
Jaynes-Cummings model with damping
\cite{tian92,rebic02,brecha99,hood-thesis}. In our paper
\cite{birnbaum05} and the accompanying Supplemental Information
\cite{birnbaum05s}, we presented theoretical results from an
extended model that described a multistate atom coupled to two
cavity modes. The relevant atomic states are the Zeeman states
of a particular hyperfine transition in atomic Cesium, namely
$6S_{1/2},F=4\leftrightarrow 6P_{3/2},F^{\prime }=5^{\prime}$ at $852$ nm.
The relevant modes of the Fabry-Perot cavity are two TEM$_{00}$ modes of the same longitudinal order but orthogonal polarizations.

Our purpose in this paper is to provide a more complete discussion of
photon blockade than in Ref.~\cite{birnbaum05}, both by way of particular
results from our model calculation and of a general framework for
characterization of photon blockade, with which we begin in Section II. In
Section III we turn to the details of our actual system and calculate the
eigenvalue structure for a two-mode cavity coupled to an atom with multiple
internal states. Here, we make explicit the coupling in the model
Hamiltonian which is used to determine the eigenvalues displayed in Figure 1(b) of Ref.~\cite{birnbaum05}. We also incorporate this Hamiltonian into the master
equation for the damped, driven system used to compute theoretical results
for transmission spectra and photon statistics, as in Figure 2(b) of Ref.~\cite{birnbaum05}. In Section IV, we gain some perspective on the relevant
physical mechanisms by comparing transmission spectra and photon statistics
for the case of an external drive that excites the atom (rather than
the cavity, as in our experiment). Section V presents an extension to our
atom-cavity model which includes the effect of cavity birefringence and FORT-induced ac-Stark shifts in the atomic states. The modified cavity transmission and intensity correlation functions
are presented for comparison to previous results. We also display a
theoretical result for the time dependence of the intensity correlation
function for comparison to our measured result in Ref.~\cite{birnbaum05}.
Finally, in Section VI we offer a discussion of these various results.

\section{General considerations}

We begin by attempting to address the question \textquotedblleft
What is photon blockade?\textquotedblright \cite{werner99}. Consider
the following input electromagnetic field state to
some \textquotedblleft black box\textquotedblright ,%
\begin{equation}
|\psi _{in}\rangle =\sum a_{n}|n\rangle \text{ .}  \label{psiin}
\end{equation}%
Assume that the mapping of input to output by the black box is given by the
transmission coefficients $t_{n}$ for each Fock-state $|n\rangle$. The output
state is then of the form%
\begin{equation}
|\psi _{out}\rangle =\sum t_{n}a_{n}|n\rangle \text{ .}  \label{psiout}
\end{equation}%
For a coherent-state input $|\alpha \rangle $, $a_{n}\varpropto \alpha ^{n}/%
\sqrt{n!}$, so that%
\begin{equation}
|\psi _{out}\rangle \propto \sum t_{n}\alpha ^{n}/\sqrt{n!}|n\rangle \text{ .%
}  \label{psiout-alpha}
\end{equation}%
A linear transfer function for the black box would be of the form $t_{n}\sim
(t_{1})^{n}$, so that%
\begin{eqnarray}
|\psi _{out}\rangle &\propto &\sum (t_{1}\alpha )^{n}/\sqrt{n!}|n\rangle \\
&\propto &|t_{1}\alpha \rangle \text{ .}  \nonumber
\end{eqnarray}

By contrast, a nonlinear device can modify the photon statistics in a
fashion other than $|t_{n}|\sim |t_{1}|^{n}$. The working criterion that we
adopt here for photon blockade is that the transmission coefficients $%
|t_{n}|<|t_{1}|^{n}$ for $n\geq 2$. For example, an \textquotedblleft
ideal\textquotedblright\ photon blockade device that eliminates all Fock states with $n\geq 2$ ($t_{n}=0$ for $%
n\geq 2$) would lead to the following output for a coherent state input:%
\begin{equation}
|\psi _{out}\rangle \propto |0\rangle +t_{1}\alpha |1\rangle \text{ .}
\label{psiout-01}
\end{equation}%
Likewise, a device that produces photon pairs in abundance (and an
associated large degree of photon bunching) could be specified by $%
|t_{n=2}|\gg |t_{1}|^{2}$, with all $t_{n>2}=0$.

Of course, even in this simple setting of $in\rightarrow out$, the above
discussion is incomplete since at least one additional input and output
channel is required to preserve unitarity. More generally, the
transformation $in\rightarrow out$ requires multiple input and output
channels (e.g., polarizations for the input and output fields with a
continuum of frequencies, relevant quantum degrees of freedom for the
material system of the black box, etc.). These additional channels may affect the coherence of the output field, as discussed below in Section VI. Within this more complex setting, however, 
the conceptual framework that we suggest for identifying photon blockade
still rests upon the simple intuition described above. Namely, one of the
output channels should have the property that the transmission coefficients $%
t_{n}$ satisfy $t_{n}<|t_{1}|^{n}$ for $n\geq 2$.

Note that the pioneering work on photon blockade based upon EIT satisfies
the above criterion \cite{imamoglu97,grangier98,werner99,rebic99,rebic02},
as does resonance fluorescence from a single atom \cite{kimble77} and the
cavity QED\ schemes considered in Refs. \cite%
{tian92,rebic02,brecha99,hood-thesis}. Indeed, by the criterion
stated above, most single atomic or molecular emitters function by
way of photon blockade. The problem of course is that the efficiency
for collecting fluorescence is typically poor. So, in addition to
the more fundamental requirement $t_{n}<|t_{1}|^{n}$, it seems
reasonable to add a second, more practical criterion related to
efficiency. Photon blockade is of much less practical significance
if the efficiency for the mapping of input to an output channel is
negligibly small, but precisely \textquotedblleft how small is too
small\textquotedblright\ is hard to quantify and depends upon the
particular application. Much of the effort related to photon
blockade is directed towards maintaining the \textquotedblleft
quality\textquotedblright\ of blockade inherent in single-atom
resonance fluorescence while at the same time achieving a sensibly
large efficiency, which has led our group to employ an optical
cavity within the setting of cavity QED, a system to which we now turn our
attention.

\section{Eigenvalues of the atom-cavity system}

We begin this section, adapted from Ref.~\cite{birnbaum05s}, by considering the eigenvalue structure of the atom-cavity
system in the absence of damping, with the model
system illustrated in Figure \ref{system}. Approximating the
atom-cavity coupling as a dipole interaction, we define the atomic
dipole transition operators for the $6S_{1/2},F=4\rightarrow
6P_{3/2},F^{\prime }=5^{\prime }$ transition in atomic Cesium as
\begin{equation}
D_{q}=\sum_{m_{F}=-4}^{4}|F=4,m_{F}\rangle \langle F=4,m_{F}|\mu
_{q}|F^{\prime }=5^{\prime },m_{F}+q\rangle \langle F^{\prime }=5^{\prime
},m_{F}+q|,  \label{block_plus_dipole}
\end{equation}%
where $q=\{-1,0,1\}$ and $\mu _{q}$ is the dipole operator for $\{\sigma
_{-},\pi ,\sigma _{+}\}$-polarization, respectively, normalized such that
for the cycling transition $\langle F=4,m_{F}=4|\mu _{1}|F^{\prime
}=5^{\prime },m_{F}=5\rangle =1$. The matrix element of the dipole operator $%
\langle F=4,m_{F}|\mu _{q}|F^{\prime }=5^{\prime },m_{F}^{\prime }\rangle $
is equal to the Clebsch-Gordan coefficient for adding spin $1$ to spin $%
4$ to reach total spin $5$, namely $\langle
j_{1}=4,j_{2}=1;m_{1}=m_{F},m_{2}=q|j_{total}=5;m_{total}=m_{F}^{\prime
}\rangle $.

\begin{table}[tbp]
\begin{tabular}{|c|l|c|l|c|}
\hline
k & $\varepsilon_{k}^{(1)}$ & $\eta_{k}^{(1)}$ & $\varepsilon_{k}^{(2)}$ & $\eta_{k}^{(2)}$ \\
 \hline 0 & 0     & 7  & 0     & 5 \\
 \hline 1 & 0.667 & 1  & 0.516 & 1 \\
 \hline 2 & 0.683 & 2  & 0.556 & 2 \\ 
 \hline 3 & 0.730 & 2  & 0.662 & 2 \\ 
 \hline 4 & 0.803 & 2  & 0.805 & 2 \\ 
 \hline 5 & 0.894 & 2  & 0.966 & 3 \\ 
 \hline 6 & 1     & 2  & 0.978 & 2 \\ 
 \hline 7 & --    & -- & 1.014 & 2 \\ 
 \hline 8 & --    & -- & 1.073 & 2 \\ 
 \hline 9 & --    & -- & 1.155 & 2 \\ 
 \hline 10 & --   & -- & 1.265 & 2 \\ 
 \hline 11 & --   & -- & 1.414 & 2 \\ \hline
\end{tabular}
\caption{Numerical factors $\protect\varepsilon _{k}^{(n)}$ for the
eigenvalues of the Hamiltonian $H_{4\rightarrow 5^{\prime }}$ in Eq.~(\protect\ref{H2}), together with their degeneracies $\protect\eta
_{k}^{(n)}$.} \label{eigenvalues}
\end{table}

The Hamiltonian of a single atom coupled to a cavity with two degenerate
orthogonal linear modes is
\begin{eqnarray}  \label{H2}
H_{4\rightarrow 5^{\prime}} &=&\hbar
\omega_{A}\sum_{m_{F}^{\prime}=-5}^{5}|F^{\prime}=5^{\prime},m_{F}^{\prime}%
\rangle \langle F^{\prime}=5^{\prime},m_{F}^{\prime}|+\hbar
\omega_{C_{1}}(a^{\dag}a+b^{\dag}b) \\
&&+\hbar g_{0}(a^{\dag }D_{0}+D_{0}^{\dag}a+b^{\dag}D_{y}+D_{y}^{\dag}b)
\nonumber
\end{eqnarray}
where $\omega_A$ is the atomic transition frequency, $\omega_{C_{1}}$ is the cavity resonance frequency, and $D_{y}=\frac{i}{\sqrt{2}}(D_{-1}+D_{+1})$ is the dipole operator for
linear polarization along the $y$-axis. We are using coordinates where the
cavity supports $\hat{y}$ and $\hat{z}$ polarizations and $\hat{x}$ is along
the cavity axis. The annihilation operator for the $\hat{z}$ ($\hat{y}$)
polarized cavity mode is $a$ ($b$).

\begin{figure}
\includegraphics[width=6in]{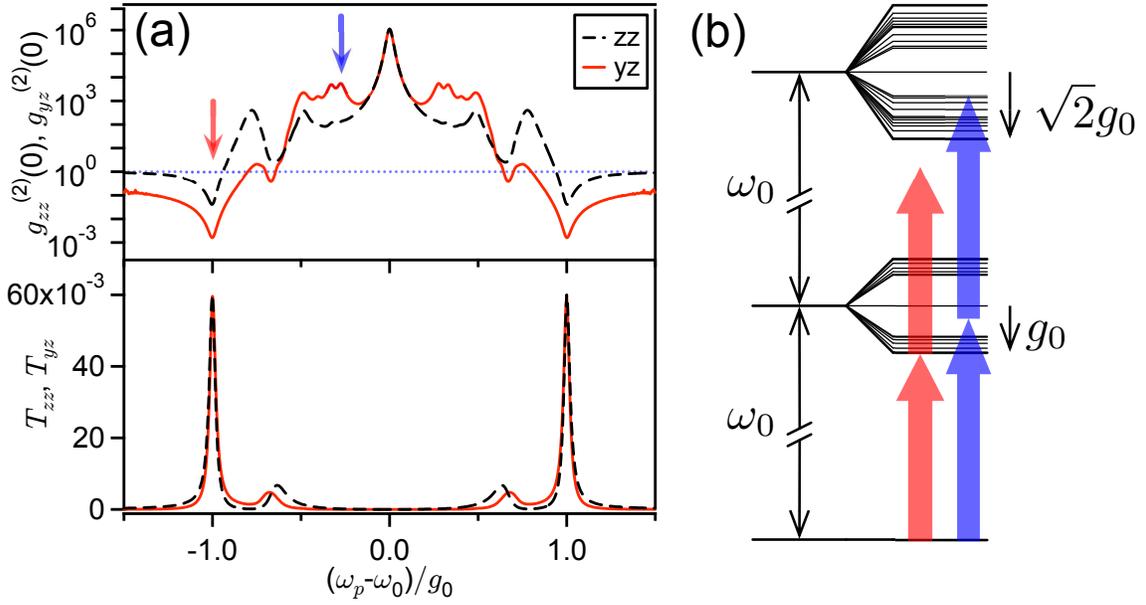}
\caption{(a) $T_{zz}$ and $g_{zz}^{(2)}(0)$ (dashed), and $T_{yz}$ and $%
g_{yz}^{(2)}(0)$ (red) versus normalized probe detuning. We consider
an $F=4\rightarrow F^{\prime }=5^{\prime }$ transition driven by
linearly polarized light in a cavity containing two modes of
orthogonal polarization that are frequency-degenerate. Parameters are $(g_{0},\protect\kappa ,\protect\gamma )/2\protect%
\pi=(50,1,1) $ MHz. The probe strength is such that the intracavity
photon number on resonance without an atom is $0.05$. The blue
dotted line indicates $g^{(2)}(0)=1$ for Poissonian statistics.  (b) Diagram of the eigenvalue structure from Table~\ref{eigenvalues}.  Red and blue arrows denote probe frequencies indicated in (a) which lead to sub-Poissonian and super-Poissonian statistics, respectively.}
\label{gbig}
\end{figure}

Assuming $\omega _{A}=\omega _{C_{1}}\equiv \omega _{0}$, we find that the
lowest eigenvalues of $H_{4\rightarrow 5^{\prime }}$ have a relatively
simple structure. In the manifold of zero excitations, all nine eigenvalues
are zero. In manifolds with $n$ excitations, the eigenvalues are of the form
$E_{n,k}=n\hbar \omega _{0}+\hbar g_{0}\varepsilon _{k}^{(n)}$, where $%
\varepsilon _{k}^{(n)}$ is a numerical factor and $k$ is an index for
distinct eigenvalues. There are $29$ states in the $n=1$ manifold, but due
to degeneracy $k$ has only $13$ distinct values, $k\in \{-6,\dots 6\}$; in
the $n=2$ manifold there are $49$ states but $k\in \{-11,\dots 11\}$. The total
number of states in any manifold can be understood by considering how the
excitations can be distributed among the atom and the two cavity modes. For
example, in the $n=1$ manifold, the atom can be in one of its $9$ ground
states ($m_{F}\in \{-4,\dots 4\}$) and either cavity mode $l_{y}$ or mode $%
l_{z}$ can have one photon (giving $18$ possible states), or the atom can be
in one of its $11$ excited states ($m_{F}^{\prime }\in \{-5,\dots 5\}$)
while both cavity modes are in the vacuum state, yielding a total of 29
states. Table \ref{eigenvalues} lists numerical values for $\varepsilon
_{k}^{(1,2)}$ as well as their respective degeneracies $\eta _{k}^{(1,2)}$.
The numerical factors and degeneracies have the symmetries $\varepsilon
_{-k}^{(n)}=-\varepsilon _{k}^{(n)}$ and $\eta _{-k}^{(n)}=\eta _{k}^{(n)}$.
The resulting eigenvalues $E_{n,k}$ for $n=\{0,1,2\}$ are displayed in Fig.~\ref{gbig}(b).

Although these eigenvalues are certainly not sufficient for understanding
the complex dynamics associated with the full master equation, they do
provide some insight into some structural aspects of the atom-cavity system.
For example, the eigenvalues $\varepsilon _{\pm 6}^{(1)}=\pm 1$ correspond
to the vacuum-Rabi splitting for the states $|1,\pm \rangle $ for a
two-state atom coupled to a single cavity mode [cf., Fig.~1(a) of Ref.~\cite%
{birnbaum05}]. The one-photon detunings for transitions from the $%
n=1\rightarrow n^{\prime }=2$ manifold are largest for the eigenstates
associated with $\varepsilon _{\pm 6}^{(1)}$. Indeed, just as for the
two-state atom with one cavity mode, transitions from the eigenstates at $%
\pm g_{0}$ have frequency detunings $\pm (2-\sqrt{2}) g_{0}$
relative to the nearest states in the $n^{\prime }=2$ manifold (at
$\varepsilon _{\pm 11}^{(2)}=\pm \sqrt{2}$, respectively). Hence, as
a function of probe frequency $\omega _{p}$, the eigenvalue
structure in Table \ref{eigenvalues} suggests that the ratio of
two-photon to one-photon excitation would exhibit a minimum around
$\omega _{p}=\omega _{0}\pm g_{0}$, resulting in reduced values
$g^{(2)}(0)<1$ \cite{g2-define}, which the full calculation verifies
in Fig.~2(b) of Ref.~\cite{birnbaum05}.

For excitation to the other eigenstates in the $n=1$ manifold, such blockade
is not evidenced in Fig.~2(b) of Ref.~\cite{birnbaum05}. A contributing
factor suggested by the structure of eigenvalues in Table \ref{eigenvalues}
is interference of one and two-photon excitation processes. For example,
excitation at $\omega _{p}\simeq\omega _{0}\pm g_{0}/4$ results in
two-photon resonance for the eigenstates associated with $\varepsilon _{\pm
1}^{(2)}\simeq \pm 0.5$, and leads to photon bunching with $g^{(2)}(0)\gg 1$ as
confirmed by our full calculation of photon statistics.

Fig.~\ref{gbig}(a) provides a global perspective of these various effects.
Here, we calculate transmission spectra and intensity correlation functions
analogous to those shown in Figure 2(b) of Ref.~\cite{birnbaum05}, but now
with coherent coupling $g_{0}$ much larger than the dissipative rates $%
(\kappa ,\gamma )$ and well beyond what we have achieved in our experiments,
$g_{0}/\kappa =g_{0}/\gamma =50$ \cite{T-define}. At $\omega _{p}=\omega _{0}\pm g_{0}$, $%
g_{yz}^{(2)}(0)\simeq 0.002$ in evidence of the previously discussed photon
blockade suggested by the eigenvalue structure in Table \ref{eigenvalues}.
As anticipated, large photon bunching results near $\omega _{p}\simeq \omega
_{0}\pm g_{0}/4$ associated with the two-photon resonance to reach the
eigenstates with $\varepsilon _{\pm 1}^{(2)}\simeq \pm 0.5$. Between these
two extremes for the eigenvalues with the largest and smallest nonzero
magnitudes ($g_{0}/4\leq |\omega_{p}-\omega_{0}| \leq g_{0}$), 
$g_{yz}^{(2)}(0)$ displays a complex structure involving multiple
excitation pathways through states in the $n=1$ manifold to reach states in
the $n^{\prime }=2$ manifold. The extremely large peak at $\omega
_{p}=\omega _{0}$ is discussed in Refs.~\cite{carmichael91,brecha99}. Similar calculations show
that the photon blockade effect described by Fig.~\ref{gbig} is
unaffected in its qualitative character if the atomic spontaneous decay rate
$\gamma $ is made much smaller than the cavity decay rate $\kappa $,
although we have not set $\gamma $ strictly to zero.

\section{Driven Atom}

Fig.~2(b) of Ref.~\cite{birnbaum05} compares the predicted photon statistics
when driving the detected cavity mode ($\hat{z}$) with the statistics when driving the other cavity mode
($\hat{y}$). The driven cavity mode has photon statistics which are
less strongly sub-Poissonian, an effect we hypothesize to be caused
by interference between the atomic dipole radiation and the coherent
drive. We will now further explore this hypothesis by considering
atom-cavity systems where the driving field is directly coupled to
the atom, instead of the cavity mode.

\begin{figure}
\includegraphics[width=5in]{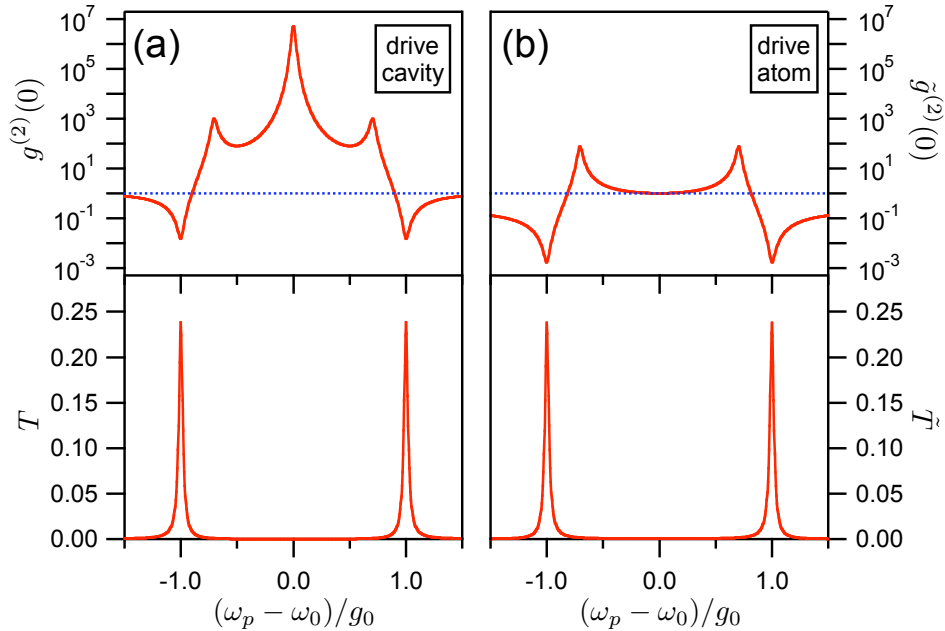}
\caption{Spectra and intensity correlation functions for the Jaynes-Cummings
system when (a) driving the cavity $T$ and $g^{(2)}(0)$ or (b) driving the
atom $\tilde{g}^{(2)}(0)$ and $\tilde{T}$ versus normalized probe detuning.
Parameters are $(g_{0},\protect\kappa ,\protect\gamma )/2\protect\pi%
=(50,1,1) $ MHz. The probe strength is such that the intracavity photon
number on resonance without an atom is $0.05$ for (a), and the atomic
excited state population on resonance without a cavity is $0.046$ ($s/2=0.05$) for (b).
The blue dotted lines indicates $g^{(2)}(0)=1$ for Poissonian statistics.}
\label{jc drivetype}
\end{figure}

We first study the familiar Jaynes-Cummings system, a two-state atom
coupled to a single mode cavity. We assume a coherent drive field
made of many photons which we will treat classically. In
Fig.~\ref{jc drivetype}, we compare the intracavity fields
\cite{carmichael85} when driving (a) the cavity and (b) the atom for
$g_{0}/\kappa = g_{0}/\gamma = 50$. $T,\tilde{T}$ are proportional
to the intracavity photon number, with $T$ normalized to
the empty-cavity on-resonance photon number for the driven cavity and $\tilde{%
T}$ normalized to half of the saturation parameter $s$ for the driven
atom. The intensity correlation function $\tilde{g}^{(2)}(0)$ of the system
when driving the atom is much lower than $g^{(2)}(0)$ (the correlation
function of the system when driving the cavity) at $\omega_p = \omega_0 \pm
g_0$, with $\tilde{g}^{(2)}(0) \simeq 0.002$ and $g^{(2)}(0) \simeq 0.02$. $%
\tilde{g}^{(2)}(0)$ is super-Poissonian at $\omega_p = \omega_0 \pm g_0/%
\sqrt{2}$ due to the two-photon resonance discussed above, but lacks the
large peak at $\omega_p = \omega_0$ evident in $g^{(2)}(0)$. At $\omega_p =
\omega_0$ we have $\tilde{g}^{(2)}(0)\simeq1$.

\begin{figure}
\includegraphics[width=5in]{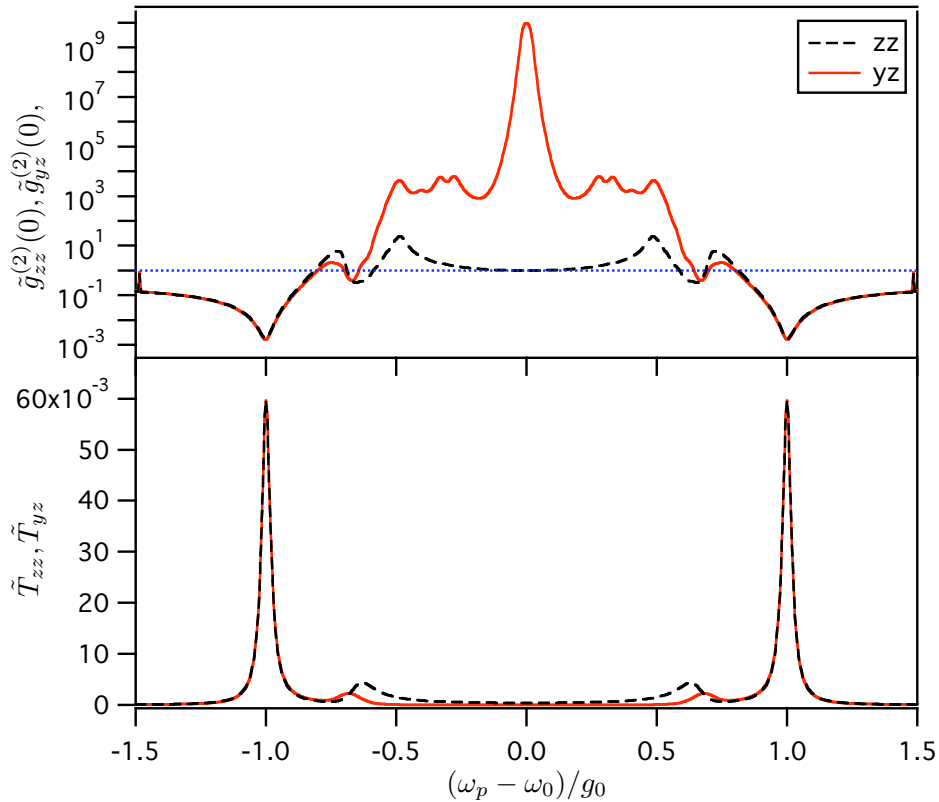}
\caption{Two-mode cavity coupled to a single Cs atom with direct
excitation of the atom. $\tilde{T}_{zz}$ and
$\tilde{g}_{zz}^{(2)}(0)$ (dashed), and $\tilde{T}_{yz}$ and
$\tilde{g}_{yz}^{(2)}(0)$ (red) versus normalized probe detuning.
Parameters are $(g_{0},\protect\kappa ,\protect\gamma )/2\protect\pi%
=(50,1,1) $ MHz. The probe strength is such that the atomic excited state
population on resonance without a cavity is $0.025$. The blue dotted line
indicates $g^{(2)}(0)=1$ for Poissonian statistics.}
\label{drive_atom}
\end{figure}

We next consider a two-mode cavity coupled to the Zeeman states of the $%
F=4\rightarrow F^{\prime }=5^{\prime }$ transition of a single atom, as
in Section III. We take the probe field driving the atom to be polarized along $%
\hat{z}$ and calculate the intracavity photon number $\tilde{T}_{zz},\tilde{T%
}_{yz}$ in the $\hat{z},\hat{y}$ modes, respectively (normalized to the
drive strength as above), and the corresponding intensity correlation
functions $\tilde{g}_{zz}^{(2)}(0),\tilde{g}_{yz}^{(2)}(0)$. Results of the
calculations are plotted in Fig.~\ref{drive_atom} for $g_{0}/\kappa
=g_{0}/\gamma =50$. At $\omega _{p}=\omega _{0}\pm g_{0}$, both $\tilde{g}%
_{zz}^{(2)}(0)$ and $\tilde{g}_{yz}^{(2)}(0)$ are close to $g_{yz}^{(2)}(0)$
as in Fig.~\ref{gbig}. This supports our hypothesis that the somewhat higher
value of $g_{zz}^{(2)}(0)$ at $\omega _{p}=\omega _{0}\pm g_{0}$ is caused
by interference with the drive field. Interestingly, though the central peak
at $\omega _{p}=\omega _{0}$ is absent in $\tilde{g}_{zz}^{(2)}(0)$, as we
expect in correspondence with the Jaynes-Cummings case and as implied by the
usual explanation of the phenomenon as resulting from the interference of
the drive field with the atomic dipole radiation, the peak in $\tilde{g}%
_{yz}^{(2)}(0)$ is even greater than that of $%
g_{zz}^{(2)}(0),g_{yz}^{(2)}(0) $. This effect may be of interest in future
studies.

\section{Birefringence and Stark Shifts}

\begin{figure}
\includegraphics[width=5in]{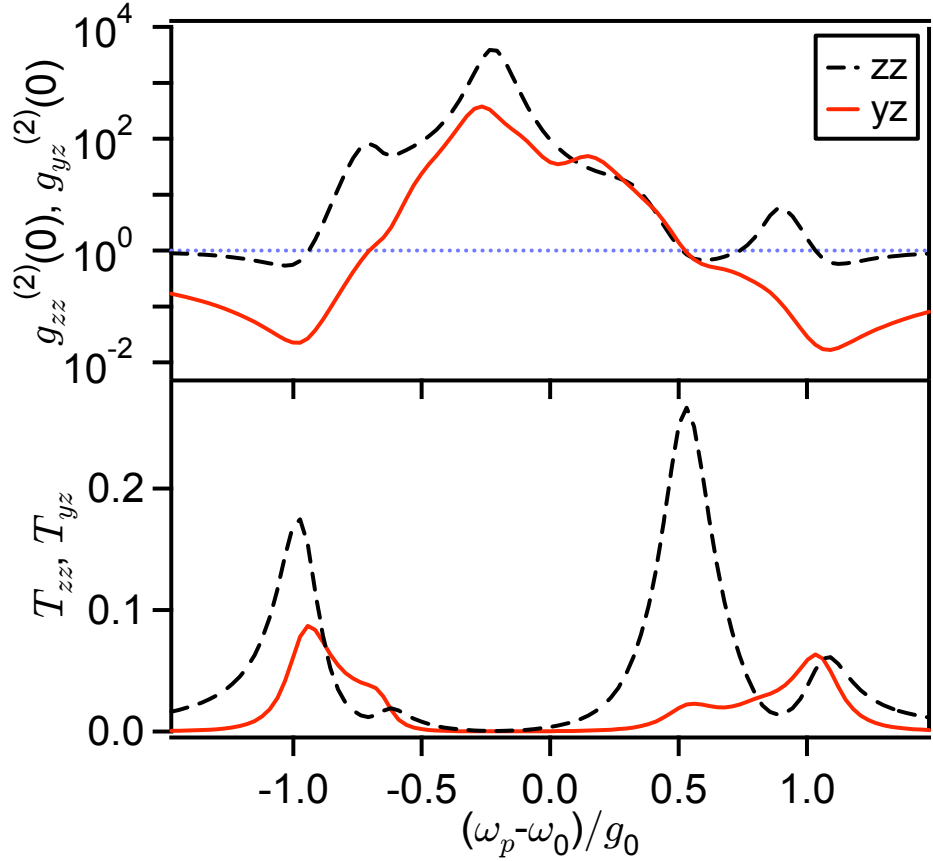}
\caption{$T_{zz}$ and $g_{zz}^{(2)}(0)$ (dashed), and $T_{yz}$ and $%
g_{yz}^{(2)}(0)$ (red) versus normalized probe detuning. We consider an $F=4\rightarrow
F^{\prime }=5^{\prime }$ transition (with FORT induced ac-Stark shifts) in a
cavity (containing two nondegenerate modes of orthogonal polarization) driven by linearly polarized light.
Parameters are $(g_{0},\protect\kappa ,\protect\gamma ,\Delta \protect\omega %
_{C_{1}},U_{0})/2\protect\pi=(33.9,4.1,2.6,4.4,-43)$ MHz, and $\protect%
\omega _{C_{1}^{z}}=\protect\omega _{A}\equiv \protect\omega _{0}$. The
probe strength is such that the intracavity photon number on resonance
without an atom is $0.05$. The blue dotted line indicates $g^{(2)}(0)=1$ for
Poissonian statistics.}
\label{starks}
\end{figure}

We now consider the effects of cavity birefringence and $m_{F}^{\prime }$%
-dependent ac-Stark shifts, expanding our previous treatment from Ref.~\cite%
{birnbaum05s}. The birefringence and ac-Stark shifts modify the Hamiltonian $%
H_{4\rightarrow 5^{\prime }}$ in Eq.~(\ref{H2}) to%
\begin{eqnarray}
H_{full} &=&\sum_{m_{F}^{\prime }=-5}^{5}\hbar \omega _{m_{F}^{\prime
}}|F^{\prime }=5^{\prime },m_{F}^{\prime }\rangle \langle F^{\prime
}=5^{\prime },m_{F}^{\prime }|+\hbar \omega _{C_{1}^{z}}a^{\dag }a+\hbar
\omega _{C_{1}^{y}}b^{\dag }b  \label{Hfulla} \\
&&+\hbar g_{0}(a^{\dag }D_{0}+D_{0}^{\dag }a+b^{\dag }D_{y}+D_{y}^{\dag }b)
\nonumber
\end{eqnarray}%
The birefringent splitting $\Delta \omega _{C_{1}}$\ is the difference of
the resonant frequencies of the two polarization modes, $\Delta \omega
_{C_{1}}=\omega _{C_{1}^{z}}-\omega _{C_{1}^{y}}$. The atomic excited state
frequencies are given by $\omega _{m_{F}^{\prime }}=\omega _{A}+U_{0}\beta
_{m_{F}^{\prime }}$, where $\omega _{A}$ is the unshifted frequency of the $%
F=4\rightarrow F^{\prime }=5^{\prime }$ transition in free space,
$U_{0}$ is the FORT potential, and $\beta _{m_{F}^{\prime }}$ for
the FORT wavelength of the experiment is given by $\{m_{F}^{\prime
},\beta _{m_{F}^{\prime }}\}$ $=$ $\{\pm 5,0.18\},$ $\{\pm
4,0.06\},$ $\{\pm 3,-0.03\},$ $\{\pm 2,-0.10\},$ $\{\pm 1,-0.14\},$
$\{0,-0.15\}$ \cite{mckeever03}.

The effect of these corrections to the Hamiltonian on the transmitted field
from the steady-state solutions to the master equation are displayed in Fig.~%
\ref{starks}, with the parameters corresponding to the experimental values from Ref.~\cite{birnbaum05}. The heights and shapes of the multiplets in
$T_{yz,zz}$ are modified, but the basic structure is unaffected
relative to Fig.~2(b) of Ref.~\cite{birnbaum05}. The structure of
$g_{yz,zz}^{(2)}(0)$ is also qualitatively unchanged. The asymmetry of the plots about $\omega_0$ is caused by the effective atom-cavity detuning (mostly due to the ac-Stark shifts). The value of
$g_{yz}^{(2)}(0)$ for $\omega
_{p}=\omega _{0}-g_{0}$ is $0.02$ (ignoring the above corrections yields $%
g_{yz}^{(2)}(0)\simeq 0.03$). These values are consistent with the
experimental result of Ref.~\cite{birnbaum05}, $g_{yz}^{(2)}(0)=0.13\pm 0.11$. Note that we
have previously reported measurements of $T_{zz}$ and made
detailed comparisons with the theory described here \cite{boca04}.

\begin{figure}
\includegraphics[width=5in]{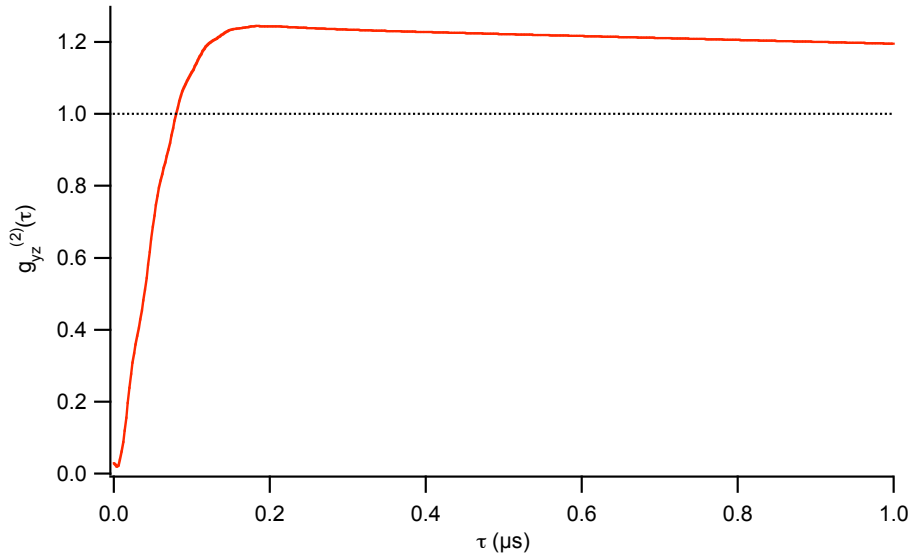}
\caption{Theoretical result for $g_{yz}^{(2)}(\protect\tau )$ versus $%
\protect\tau $ with parameters as in Fig.~\protect\ref{starks}. The probe
strength is such that the resonant intracavity photon number of the bare
cavity would be $0.21$ as in the experiment of Ref.~\cite{birnbaum05}. The blue dotted line indicates $%
g^{(2)}(0)=1$ for Poissonian statistics. This calculation was performed with
a Fock basis of $\{0,1,2\}$ photons in the $\hat{z}$ polarized mode and $%
\{0,1\}$ photons in the $\hat{y}$ polarized mode.}
\label{g2tau}
\end{figure}

In Fig.~\ref{g2tau}, we present the theoretical prediction for $%
g_{yz}^{(2)}(\tau )$ including the effects of Stark shifts and
birefringence. We find that $g_{yz}^{(2)}(\tau )$ rises to unity at
$\tau \simeq 80$ ns, well above the experimental result of $\tau
\simeq 45$ ns. The theory, however, does not take into account
atomic motion, but rather assumes the atom to be fixed at a place of
optimal coupling. Since atomic motion clearly has a large effect on
$g_{yz}^{(2)}(\tau )$, as is evident in Fig.~4(b) of Ref.~\cite{birnbaum05}, we believe that this is the dominant cause of the
discrepancy \cite{diedrich87}.

\section{Discussion}

From the point of view described in general terms in Section II and
elaborated in more detail in Sections III-V, our work in Ref.~\cite%
{birnbaum05} satisfies the criteria for photon blockade. As
indicated by Fig.~1(a,b) of Ref.~\cite{birnbaum05}, the
Jaynes-Cummings ladder provides a means to achieve the condition
$t_{2}<|t_{1}|^{2}$, with the consequence that the intensity
correlation function $g^{(2)}(0)<1$ as presented in Fig.~2 of 
Ref.~\cite{birnbaum05} and in the previous sections. Note also that by
tuning to a two-photon resonance, our calculations indicate that
photon bunching $g^{(2)}(0)\gg 1$\ could be achieved for our atom-cavity system with ($\omega
_{p}\simeq \omega _{0}\pm g_{0}/4$), again in accord with an
understanding based upon transmission coefficients $t_{n}$. Photon
bunching for the standard Jaynes-Cummings ladder of Fig.~1(a) in Ref.~\cite%
{birnbaum05} is likewise achieved by tuning to the two-photon
resonance at $\omega _{p}\simeq \omega _{0}\pm g_{0}/\sqrt{2}$.

Our criteria for photon blockade do not demand the preservation of
coherence in the transformation from input to output. One way to express a
requirement for coherence in terms of the generic model in Section II is
that the output state should be of the form of Eq.~(\ref{psiout-01}), and not
of the form%
\begin{equation}
\rho _{out}\sim |0\rangle \langle 0|+|t_{1}\alpha |^{2}|1\rangle \langle 1|%
\text{ ,}  \label{rhoout-01}
\end{equation}%
with the coherent amplitude lost. However, our view is that either Eq~(\ref%
{psiout-01}) or Eq.~(\ref{rhoout-01}) suffices and qualifies as photon
blockade. In fact, in practice the latter case of Eq.~(\ref{rhoout-01}) might
be the more \textquotedblleft useful\textquotedblright\ for the following
reasons.

With reference to Fig.~8 in Ref.~\cite{rebic02}, note that there are various
contributions to $g^{(2)}(0)$ as a function of the amplitude of the driving
field [Eq.~(11b) of Ref.~\cite{rebic02}]. The authors point out that
\textquotedblleft The decomposition shows how the behavior of $g^{(2)}(0)$ \ldots\ can be interpreted as the effect of self-homodyning between the
coherent and incoherent components of the intracavity
field.\textquotedblright\ For our initial experiments with relatively modest
ratios $g/(\kappa ,\gamma )$, our view is that this complex, phase-dependent
interplay should be avoided since it makes the blockade effect more
\textquotedblleft fragile\textquotedblright\ (less robust) than is the case
for Eq.~(\ref{rhoout-01}). Basically, one wants a situation where there is no
need to balance a set of interference terms (as in Eq.~(11b) of Ref.~\cite%
{rebic02}), but rather a more \textquotedblleft generic\textquotedblright\
requirement of the sort presented in Section II, namely $|t_{n}|<|t_{1}|^{n}$.
 Note that an interpretation of photon antibunching similar to that
expressed in Ref.~\cite{rebic02} can be given for single-atom
resonance fluorescence, as was first analyzed by Carmichael
\cite{carmichael85}. However, in this case the \textquotedblleft
miracle\textquotedblright\ is that the terms always sum to give
$g^{(2)}(0)=0$ for any drive strength, which is not the case for an
atom in a cavity \cite{carmichael85}.

Our calculations show that the mean value for the amplitude of the
transmitted field with polarization orthogonal to that of the coherent state
input is zero. So, there is not a sense in which the coherent amplitude of
the input is rotated in polarization by the atom-cavity system in the
transformation to the output. Nevertheless, the orthogonally polarized output
field can exhibit quantum interference, with an example being the value $%
g_{yz}^{(2)}(0)\gg 1$ for $\omega _{p}=\omega _{0}$ in Figs. \ref{gbig}, \ref%
{starks}, which is presumably associated with the quantum state reduction
and interference described by Carmichael and coworkers \cite%
{carmichael91,brecha99}, with here $g_{yz}^{(2)}(0)\simeq g_{zz}^{(2)}(0)$
for $\omega _{p}=\omega _{0}$.

For our experiment with multiple Zeeman states in the ground and
excited levels and with two orthogonally polarized cavity modes
$(y,z)$, the eigenstates that are being driven for excitation along
$y$ with $\omega _{p}=\omega _{0}\pm g_{0}$ are complex
superpositions of various atomic Zeeman states and field states for
the $(y,z)$ polarizations. These eigenstates are entangled, so it is
perhaps not surprising that examination of a particular component
(e.g., for the field or atomic coherences) results in a mixed state
with mean zero. Hence, many coherences for a single degree of
freedom may vanish not because of dissipation \textit{per se}, but
rather because of entanglement with other components. 
Although we have not explored this
question in detail, we think that this may explain why the mean
amplitude for the transmitted $z$ field is zero for excitation along
$y$, as is the case for our results in Figs. \ref{gbig},
\ref{starks}.

A final perspective to offer is to extend the discussion from the
case with continuous excitation as has been implicit above to the
case of pulsed excitation, as in Fig.~4 of Ref.~\cite{imamoglu97}.
Assume a pulse with duration short compared to any time scale
associated with the \textquotedblleft black box\textquotedblright\
described in Section II.\ For resonance fluorescence, consider a
$\pi $ pulse with duration $\tau _{p}\ll \gamma ^{-1}$, so that
there is now no coherent component for the fluorescent light for times $t>$ $%
\tau _{p}$. In this case, resonance fluorescence would no longer
satisfy the criterion for the preservation of coherence (as is the
case for weak cw excitation \cite{mandel-wolf95}), yet it would
still be perfectly
antibunched. For the case of our atom-cavity system, we would require $%
g_{0}^{-1}\ll \tau _{p}\ll \kappa ^{-1}$, in which case we would presumably
obtain single photons on a pulse-by-pulse basis for the transmitted field
with polarization orthogonal to that of the drive field, again with no
preservation of coherence, which is presumably also true for EIT\ schemes in
the limit of short pulses for the excitation \cite%
{imamoglu97,grangier98,werner99,rebic99,rebic02}. Note that in
either of these cases, the efficiency for the transformation of the
pulsed driving field to a single photon at the output is necessarily
small because of the mismatch of bandwidths, $\tau _{p}\ll \kappa
^{-1}$.

\section{Acknowledgements}

We gratefully acknowlege stimulating discussions with Atac Imamo\={g}lu upon
which the material in Sections II and VI is based. This research is
supported by the National Science Foundation, by the Caltech MURI Center for
Quantum Networks, and by the Advanced Research and Development Activity
(ARDA).


\begin{thebibliography}{99}
\bibitem{T-define} $T(\omega_{p})$ is proportional to the ratio of photon
flux $\langle \mathcal{E}_{t}^{\dagger }\mathcal{E}_{t}\rangle $\
transmitted by $M_{2}$ to the flux $|\mathcal{E}_{p}|^{2}$ incident upon $%
M_{1}$, and normalized such that a cavity without an atom has a
resonant transmission of unity.  $T_{zz}$ ($T_{yz}$) is proportional to the transmitted field polarized along $\hat{z}$ for an incident probe field polarized along $\hat{z}$ ($\hat{y}$).  $g^{(2)}_{zz}$ ($g^{(2)}_{yz}$) is the intensity correlation function of the transmitted field polarized along $\hat{z}$ for an incident probe field polarized along $\hat{z}$ ($\hat{y}$).
\end{thebibliography}
\end{document}